\documentclass{sig-alternate}

\usepackage[english]{babel}
\usepackage{blindtext}
\usepackage{etoolbox}
\makeatletter
\patchcmd{\maketitle}{\@copyrightspace}{}{}{}
\makeatother

\usepackage{amsfonts}
\usepackage{amsmath}
\usepackage{graphicx}
\usepackage{hyperref}

\begin{document}
%
\conferenceinfo{CIKM'11,}{October 24--28, 2011, Glasgow, Scotland, UK}

\title{{TopSig}: Topology Preserving Document Signatures}
%
%
%
%
%

\numberofauthors{2} 
%
\author{
%
%
\alignauthor
Shlomo Geva\\
       \affaddr{Faculty of Science and Technology}\\
       \affaddr{Queensland University of Technology}\\
       \affaddr{Brisbane, Australia}\\
       \email{s.geva@qut.edu.au}
\alignauthor
Christopher M. De Vries\\
       \affaddr{Faculty of Science and Technology}\\
       \affaddr{Queensland University of Technology}\\
       \affaddr{Brisbane, Australia}\\
       \email{chris@de-vries.id.au}
}

\maketitle
\begin{abstract}
Performance comparisons between File Signatures and Inverted Files for text 
retrieval have previously shown several significant shortcomings of file
signatures relative to inverted files. The inverted file approach underpins most
state-of-the-art search engine algorithms, such as Language and Probabilistic
models. It has been widely accepted that traditional file signatures are
inferior alternatives to inverted files.  This paper describes TopSig, a new
approach to the construction of file signatures.    Many advances in semantic
hashing and dimensionality reduction have been made in recent times, but these
were not so far linked to general purpose,  signature file based, search
engines. This paper introduces a different signature file approach that builds
upon and extends these recent advances.  We are able to demonstrate significant
improvements in the performance of signature file based indexing and retrieval,
performance that is comparable to that of state of the art inverted file based
systems, including Language models and BM25.  These findings suggest that file
signatures offer a viable alternative to inverted files in suitable settings and
from the theoretical perspective it positions the file signatures model in the
class of Vector Space retrieval models.  
\end{abstract}

\category{H.3.3}{Information Storage and Retrieval}
{Information Search and Retrieval}
[Retrieval Models, Relevance Feedback, Search Process, Clustering]

\terms{Algorithms, Experimentation, Performance, Theory}

\keywords{Signature Files, Random Indexing, Topology, Quantisation, Vector
Space IR, Search Engines, Document Clustering, Document Signatures}

\vspace{1em}

\section{Introduction}
Document signatures have been largely absent from mainstream IR publications
about \em general-purpose \em search engines and ranking models for several
years. The decline in the attention paid to this approach, which had received a
lot of attention earlier, started with the publication of the paper ``Inverted
Files Versus Signature Files for Text Indexing'' by Zobel et al
\cite{Zobel1998}. This paper offers an extensive comparison between Signature
Files and Inverted Files for text indexing.  The authors have systematically and
comprehensively evaluated Signature files and  Inverted File approaches. Having
examined several general approaches they concluded that inverted files are
distinctly superior to signature files.  Signature files are found, in their
studies, to be slower, to offer less functionality, and to require larger
indexes.  They conclude that the Bit Sliced signature files under-perform on
almost all counts and offer very little if any advantages over inverted files.
Further discussion of signature files is offered in \cite{Witten1999}, and a
similar picture emerges there too. It is clear from the experimental evidence
that Bit Sliced signatures are not able to compete with state of the art
inverted file approaches in terms of retrieval performance.  Furthermore, the
presumed advantages of efficient bit-wise processing and the potential for index
compression are not generally achievable in practice.  Signature files are found
to be larger than inverted file indexes. This is perhaps surprising because
Signature files were largely motivated by the desire to represent entire
documents as relatively short bit strings, and having fixed the signature
length, the document signature is independent of actual document length.  As it
turns out, it is not possible to achieve competitive performance goals with
compact signatures and consequently signatures require even more space than
compressed indexes.  

For the sake of completeness, and since we offer a radically different approach
to the construction of file signatures, it is necessary to describe the
conventional approach first.  Conventional Bit Sliced signature files, as
described in \cite{Faloutsos1984} exploit efficient bit-wise operators that are
available on standard digital processors.  Unlike probabilistic models of IR and
Language Models, the Signature File approach is presented in an ad-hoc manner
and is computationally motivated by efficient bit-wise processing, and without
specific grounding in Information Retrieval theory.  In traditional signature
files a document is allocated a fixed-size signature of $N$ bits. Each term that
appears in the collection is assigned a random signature of width $N$, where
only a small number of $n<<N$ bits are set to $1$ with the use of a suitable
hash function.  Naturally, term signatures tend to collide on some bit
positions, but this is of course unavoidable unless the number $N$ is extremely
large, and the number $n$ is very small.  Given that the vocabulary of a
document collection typically contains millions of distinct terms, collisions
will occur, and frequently.  The document signature in this approach is derived
as the conjunction of all the term signatures within the document (\em bit-wise
\em ORed). Query terms are similarly assigned a signature.  A query is then
evaluated by comparing the query signature to each document signature.  Any
document whose signature matches every bit that is set in a query term, is taken
as a potential match. It is only a potential match because hash collisions in
generating term signatures can lead to \em false matches \em -- situations where
all the bits match, but the actual query term is not present in the document.
Consequently, documents are retrieved and checked directly against the
query to eliminate false matches.  This is a very expensive operation even if
the collection fits in memory, but with a disk based collection -- the most
likely scenario -- this is prohibitively expensive.

Indeed, the method used in traditional file signatures is known in other
domains as a Bloom Filter. B.H. Bloom in first described bloom filters
1970 \cite{Bloom1970}, and this well predates file signatures.

It is clear from the experimental evidence that Zobel et al \cite{Zobel1998}
and Witten et al \cite{Witten1999} provide, that such signatures are not likely
to compete with state of the art inverted file approaches in terms of retrieval
performance.  Furthermore, the presumed advantages of efficient bit-wise
processing and the potential for index compression are not generally achieved in
practice.  Signature files are found to be larger than inverted file indexes.

Recent approaches to similarity search \cite{Zhang2010} have explored similar
ideas to TopSig for mapping documents to $N$ bit strings for comparison using
Hamming distance. The approach taken by Zhang et al \cite{Zhang2010} and prior
publications focus on similarity comparisons between documents. Their models
have not been applied to general-purpose ad-hoc retrieval. More importantly,
Zhang et al \cite{Zhang2010} use a complicated approach to the static, off-line
derivation of signatures, and which involves supervised and unsupervised
learning to generate document signatures.  This in effect prevents the
application of the approach to ad-hoc retrieval where a query signature has to
be derived at run-time. It is not practical in a very large collection due to
the excessive computational load of supervised and unsupervised learning.

Unlike earlier attempts, we approach the design of TopSig document signatures
from basic principles. TopSig is radically different from a Bloom filter in the
construction of file signatures and in the manner in which the search is
performed.  We present results of extensive experiments, performed with large
standard IR collections, where we compare TopSig with standard retrieval models
such as BM25 and various Language Models.  We also describe document clustering
experiments that demonstrate the effectiveness of the approach relative to
standard document representation for clustering.

The remainder of this paper is organised as follows.  Section~\ref{sec:topsig}
introduces the TopSig approach in detail. Sections~\ref{sec:adhoc},
\ref{sec:eval_results} and \ref{sec:clustering} define and evaluate the use of
this approach for ad-hoc retrieval and clustering. The paper is concluded with a
discussion in Section~\ref{sec:discussion}.

\section{TopSig}
\label{sec:topsig}
TopSig represents a radically different approach to the construction of
signature files.  Unlike the traditional ad-hoc approach \cite{Faloutsos1984},
TopSig is principled and signature files emerge naturally from a highly
effective compression of the well understood and commonly used Vector Space
representation of documents.   

We approach the design of document signatures from the perspective of
dimensionality reduction. TopSig starts from a straight forward application of a
vector space representation of the collection -- the term-by-document weight
matrix. We then derive the signatures through extreme and lossy compression, in
two steps, to produce topology preserving binary document signatures.  While the
actual mechanism that is proposed is highly efficient in signature construction
and in searching, we first focus the discussion on the conceptual approach, its
justification and theoretical grounding, while leaving the implementation and
performance analysis details for later in the paper.

In this section we describe the concepts that underpin TopSig. These concepts
are not new -- Random Indexing and Numeric Quantisation -- but when put together
to form file signatures, the results are remarkable.

\subsection{Random Indexing vs LSA}
\label{sec:Random Indexing}
Latent Semantic Analysis \cite{Deerwester1990} is a popular technique that is
used with word space models. LSA \cite{Golub1965} creates context vectors from a
document-by-term occurrence matrix by performing Singular Value Decomposition
(SVD).  Dimensionality reduction is achieved through projection of the
document-by-term occurrence vectors onto the subspace spanned by relatively few
vectors having the largest singular values in the decomposition.  This
projection is optimal in the sense that it minimises the Frobenius norm of the
difference between the original and the projected matrix. SVD is computationally
expensive and this limits its application in large collections. For instance, in
our own experiments, the SVD of a collection of 25,000 English Wikipedia
articles -- less than 1\% of the collection -- using the highly efficient
parallel multi-processor implementation of the MATLAB svds function, took about
7 hours on a top-end quad-processor workstation with sufficient memory to be
completely processor bound.

Random Indexing (RI) \cite{Sahlgren2005} is an efficient, scalable and
incremental approach to dimensionality reduction. Word space models often use
Random Indexing as an alternative to Latent Semantic Analysis.  Both LSA and RI
start from the term-by-document frequency matrix.  Often term frequencies are
replaced by term weights -- for instance, one of the many TF-IDF variants.
With LSA, Singular Value Decomposition is used to derive an optimal projection
onto a lower dimensional space. Random Indexing is based on a random projection
- avoiding the computational cost of matrix factorisation.  Having obtained a
projection matrix, both LSA and RI proceed to project the term occurrence matrix
onto a subspace of significantly reduced dimensionality.

In practice, RI works with one document at a time, and one term at a time within
the document. Terms are assigned random vectors, and the projected document
vector is then the arithmetic sum of all term signatures within.  The process is
somewhat similar to the traditional signature file approach of
\cite{Faloutsos1984}, but the document vector is \em real valued\em; it is a
superposition of all the random term vectors. There is no matrix factorisation
and hence the process is efficient. It has linear complexity in the number of
terms in a document and also in the collection size. This is a significant
advantage over LSA whose time complexity is prohibitive in large collections. As
stated by Manning et al \cite{Manning2008} in 2008, in relation to LSA --  \em
``The computational cost of the SVD is significant; at the time of this writing,
we know of no successful experiment with over one million documents''. \em

The RI process is conceptually very different from LSA and does not carry the
same optimality guarantees.  At the foundation of RI is the
Johnson-Lindenstrauss lemma \cite{Johnson1984}. It states that if points in a
high-dimensional space are projected into a randomly chosen subspace, of
sufficiently high-dimensionality, then the distances between the points are
approximately preserved.   Although strictly speaking an orthogonal projection
is ideal, nearly orthogonal vectors can be used and have been found to perform
similarly \cite{Bingham2001}. These vectors are usually drawn from a random
uniform distribution.  This property of preserving relative distances between
points is useful when comparing documents in the reduced space. RI offers
dimensionality reduction at low computational cost and complexity while still
preserving the topological relationships amongst document vectors under the
projection.

In RI, each dimension in the original space is given a randomly generated index
vector. The index vectors are high dimensional, sparse, and ternary. Sparseness
is controlled via a parameter that specifies the number of randomly selected
non-zero dimensions. Ternary term vectors consist of randomly and sparsely
distributed +1 and -1 values in a vector that otherwise consists mostly of
zeros.  This choice ensures that the random vectors are near orthogonal.

RI can be expressed as a matrix multiplication of a randomly generated
term-by-signature matrix $T$ by a term-by-document matrix $D$ where $R$ is the
randomly projected term-by-document matrix.

\begin{equation}
R_{N \times d} = T_{N \times t}D_{t \times d}
\end{equation}

Each of the $d$ column vectors in $D$ represents a document of dimensionality
$t$, each of the $t$ column vectors in $T$ is a randomly generated term vector
of dimensionality $N$.  $R$ is the reduced matrix where each of the $d$ column
vectors represents a randomly projected document vector of dimensionality $N$.

RI has several advantages over LSA. It can be performed incrementally and online
as data arrives. Any document can be indexed independently from all other
documents in the collection.  This eliminates the need to build and store the
entire term-by-document matrix. Additionally, newly encountered terms are
naturally accommodated without having to recalculate any of the projections of
previously encoded documents.  By contrast, LSA requires global analysis where
the number of documents and terms are fixed. The time complexity of RI is also
very attractive. It is linear in the number of terms in a document and hence
linear in the collection size.  RI makes virtually no demands on computer memory
since each document is indexed in turn and the signatures are independent of
each other.

TopSig deviates from Sahlgren's basic Random Indexing by introducing term
weights into the projection. In Sahlgren's scheme, the term-by-document matrix
contains unweighted term counts. Search engine evaluation consistently shows
that unweighted term frequencies do not produce the best performance. Better
results are obtained if the terms frequencies are weighted and this of course
underlies the most successful search engine models, such as BM25 and Language
models.  The weighting of terms in TopSig is described in
Section~\ref{sec:TopSig}.

Term weighting has an apparent drawback -- it may appear to compromise the
ability to encode new documents independently. The calculation of term weights,
such as with TF-IDF or Language Models, requires global statistics. We observe
however that in a large collection new documents have very little impact on
these global statistics. Upon inserting a new document these global statistics
are updated and the new document is encoded. As the collection grows, it is
periodically re-indexed from scratch to bring all signatures into line, but this
is a relatively cheap operation. On a modern multi-processor PC using the
ATIRE search engine \cite{Trotman2010} we can index the entire English Wikipedia
of 2.7 million documents, spanning 50 gigabytes of XML documents, in under 15
minutes.

\subsection{Random Indexing and Other Approaches}
Random Indexing shares many properties with other approaches. In this section
we will highlight some of the more interesting properties shared with other
dimensionality reduction approaches.

RI or random projections are closely related to compressed sensing from the
field of signal processing. Compressed sensing is able to reconstruct
signals with less samples than required by the Nyquist rate. Baraniuk et al
\cite{Baraniuk2008} construct a proof showing how the Restricted Isometry
Property that underlies compressed sensing is linked to the
Johnson-\\Lindenstrauss lemma \cite{Johnson1984} which underlies RI.

A conceptually similar approach to RI is used for a spread spectrum approach in
Carrier Division Multiple Access \cite{Rappaport1996}. In contrast,
CDMA uses orthogonal vectors for codes and increases the bandwidth of the
signal. In CDMA, the use of random orthogonal codes allows for division of the
radio spectrum that is more resistant to noise introduced in radio frequency
transmission.

Many other approaches to dimensionality reduction exist. Again, many come from
the field of signal processing. Many of these approaches iteratively optimise
an objective function. LSA offers an optimal linear projection in preserving the
Frobenius norm. Other well known approaches include the Discrete Cosine
Transform, Wavelet Transform, Non-Negative Matrix Factorisation, Principal
Component Analysis and Cluster Analysis. The advantage to RI is that it still
preserves the topological relationships between the vectors without having to
directly optimise an objective function. This is where its computational
efficiency comes from.

\subsection{TopSig Signatures}
\label{sec:TopSig}
Document Signatures are fixed length bit patterns. In order to transform the
real-valued projected term-by-document matrix into a signatures matrix, we ask
what
numerical precision is required to represent the term-by-document matrix.  It is
obvious that there is no need for double precision and one obtains identical
results when evaluating searching or clustering performance with single
precision.  One quickly finds that even when scaling the values to the range
[0,255] -- i.e. a single byte -- there is no appreciable difference. Even
Nibbles (4-bit integers) have been shown to be sufficient with little
appreciable difference in performance.  This is exploited by all state of the
art search engines to compress indexes.  The reduction in precision still leaves
the term-by-document matrix with a highly faithful representation of the
similarity
relationships between the original documents.  Both clustering and ranking
applications are concerned not with the actual similarity values, but rather
with their rank order.  As long as rank order is preserved the distortion due to
reduced numerical precision is not problematic.

In section \ref{sec:Random Indexing} we described how a real-valued document
vector is obtained through random projection, as the sum of random term
signatures within.  TopSig now takes the reduction in numerical precision to its
ultimate conclusion, by taking this real-valued randomly indexed document, and
reducing the precision all the way to a single bit. Binary signatures are
obtained by taking only the \em sign-bits \em of the projected document vectors
(!!).  This is a key step in TopSig signature calculation; it may appear to be
highly excessive precision reduction, but it is in fact surprisingly effective,
as we shall demonstrate with search and clustering experiments in the following
sections.

\subsubsection{Topological Distortion}
In order to measure the impact of aggressive dimensionality reduction we
conduct the following experiment.  We take 1000 randomly chosen Wikipedia
document vectors, in full TF-IDF representation, and compute their mutual
distance matrix.  Each element in the matrix represents the distance between a
pair of document vectors in the full space.  We then randomly project the
vectors onto a lower dimensional subspace and compute the corresponding mutual
distance matrix in the projection subspace.  The mutual distance matrices are
normalised such that the sum of elements in each matrix is equal to 1.  If the
mutual distances are perfectly preserved then the normalised matrices will be
identical. However, with aggressive compression we expect a topological
distortion due to information loss. To measure the impact, we calculate the
topological distortion as  the root mean squared differences (RMSE) between
distances in full precision, and the corresponding distances in the reduced
dimensionality and reduced precision.  This calculation is performed for various
dimensionality reduction values and various numeric precision values. 

\begin{figure}
\includegraphics[width=\columnwidth]{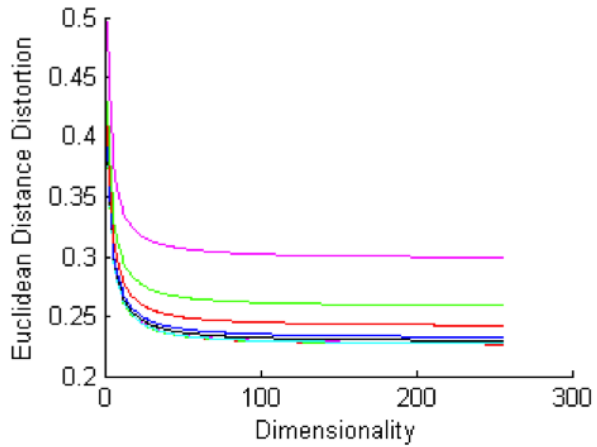}
\caption{RMSE Drop Precision}
\label{fig:rmse_precision}
\end{figure}

Figure~\ref{fig:rmse_precision} depicts the results of our experiment. On the
y-axis is the topological distortion, measured by RMSE. On the x-axis is the
number of dimensions in the projection.  Each of the curves on the plot
corresponds to a different numerical precision.  The bottom curve corresponds to
double precision, and then the plots above correspond to 8-bit quantisation,
through 7-bit quantisation, and so on all the way down to 1-bit quantisation.
First we observe that as the dimensionality of the projected subspace is
increased (moving to the right with the curves), the distortion becomes smaller.
 This is true regardless of numerical precision and it is expected. We also
observe that most of the gain is achieved quite early with relatively small
dimensionality.  This is the expected behaviour of both RI and LSA, where a
relatively small number of dimensions typically is required to achieve good
results with text documents.  What is perhaps less expected is that as we reduce
the numerical precision the deterioration is very small.  The lowest curve in
Figure~\ref{fig:rmse_precision}  corresponds to double precision.  It is only
when precision is dropped to 3-bit that the difference in RMSE becomes
noticeable.   The curves from 8-bit down to to 4-bit quantisation are barely
separated.    The distortion only increases significantly when we drop to 3-bit,
2-bit and 1-bit precision, corresponding to the 3 higher curves in the figure.
Even with 1-bit precision we are still able to significantly preserve topology
quite early with very small dimensionality. 

\subsubsection{Packing Ternary Vectors onto Binary Strings}
\label{sec:packing}
To complete the generation of a document signature we need to pack the $\pm1$
representation of signatures, onto binary strings.  This is done by representing
positive signs as 1s, and negative signs as 0s. The final result is thus a
binary digital signature, but it still conceptually represents  $\pm 1$
signatures.  

We note that there is a possibility that very short documents will not occupy
all bit positions in a signature.  We can safely ignore this situation and
encode zeros as positive (i.e. binary $1$) although it may introduce some noise
into the representation.  The effect is negligible and studying it is outside
the scope of this paper.  Suffice to say that in our experiments circumventing
this by complicating the representation to also record the unoccupied positions
resulted in no appreciable difference at all and there was no practical
advantage to maintaining this information. 

\subsubsection{Summary of Binary Signatures}
To summarise, TopSig introduces a principled approach to the generation of
binary file signatures. The underlying data representation starts exactly as
with inverted files, from the term-by-document weight matrix.  This matrix is
then subjected to aggressive lossy compression.   Topology preserving
dimensionality reduction is first achieved through Random Indexing and it is
immediately followed by aggressive numerical precision reduction by keeping only
the sign bits of the projected term-by-document weight matrix.  Unlike
traditional signatures, TopSig does not emerge from bit-wise processor
efficiency considerations, but rather, it emerges as a consequence of aggressive
compression of a well understood document representation.  In this scheme,
document signatures are no more than highly concise approximations of vector
space document representations.  TopSig maps an entire document collection onto
corners of the $\{\pm 1\}^N$ hypercube.

\section{Ad-hoc Retrieval}
\label{sec:adhoc}
To provide a concrete description of the implementation and use of TopSig in
ad-hoc retrieval we need to more precisely define document and query signatures,
term weights, the ranking process, and how pseudo-relevance feedback is used. We
then describe the evaluation of document retrieval using the INEX Wikipedia
collection and the TREC Wall Street Journal (WSJ) collection.  We conclude this
section with the description of document clustering experiments.

\subsection{Document Signatures}
\label{sec:docsig}

So far we have not addressed the weighting of terms in the vector space
representation of the term-by-document matrix.  Weighting is all-important to
improving precision and recall and it is the basis of the most successful
ranking functions, such as BM25 and Language models, which compute term weights
in many different ways.   With TopSig, the most effective weighting function we
have found is described in
Equations~\eqref{eq:td},~\eqref{eq:tc}~and~\eqref{eq:tftdtc}

\begin{equation}
\label{eq:td}
P(t|D) = \frac{tdf}{|D|}
\end{equation}

\begin{equation}
\label{eq:tc}
P(t|C) = \frac{tcf}{|C|}
\end{equation}

\begin{equation}
\label{eq:tftdtc}
W(t,D) = log\left(\frac{P(t|D)}{P(t|C)}\right)
\end{equation}

\noindent where $W(t,D)$ is the weight for term $t$ in document $D$. We define
$tdf$ to be the term frequency for term $t$ in document $D$, $|D|$ as the total
number of term occurrences in document $D$, $tcf$ as the collection frequency
for term $t$, and $|C|$ as the total number of term occurrences in the
collection. $P(t|D)$ is an estimate of the probability of finding the term $t$
given a document $D$, and $P(t|C)$ is an estimate of the probability of finding
term $t$ given the collection $C$. 

The weighting function $W(t,D)$ produces a larger value if the frequency of a
term in a document is higher than expected, and smaller if the frequency is
lower than expected.  The logarithm of the ratio of these expected values is
taken, so as to dampen the effect of an inordinately large frequency of a term
in a document. 

The representation of a document is thus a bag of words, where the weight
assigns an individual importance score to each term within a document.  This
effectively takes care of stop-words.  We note that a term that occurs with
approximately the expected frequency will have a weight close to zero.  Negative
weights that result from equation \ref{eq:tftdtc} are set to zero since that
would indicate that the term occurs in the document with even lower frequency
than expected.  This weighting scheme ensures that stop words are naturally
discounted without special treatment. Anecdotally, a document that consists of
only the sentence ``To be, or not to be, that is the question" will retain all
terms with appreciable weights when generating this document's signature, but
most terms will have virtually negligible weights in much larger documents and
thus the terms will be effectively stopped.

\subsection{Alternative Term Weighting Functions}
Surprisingly, TopSig performs quite respectably with no term weighting at all.
The raw unweighted term frequencies and simply randomly indexed. One advantage
of this approach is that is requires no global statistics at all -- a document
can be encoded purely by looking at the document in isolation.

When using the BM25 weighting function to create a vector space representation,
we found the retrieval performance was relatively very poor. This is not
surprising as BM25 was originally intended to be treated as a probabilistic
model and we did not use it in that manner.

The TF-IDF representation produces retrieval quality that lies between raw term
frequencies and the approach described in Section \ref{sec:docsig}.

The detailed comparison of different weighting functions is outside the scope
of this paper. What we provide here is anecdotal evidence to paint a clearer
picture of the approach we have taken to developing TopSig.

\subsection{Alternative Document Representations}
While the representation we have described here for ad-hoc retrieval is a bag
of words model for keyword search, there is no limitation of encoding other
representations using the TopSig approach. For example, it is possible to
create vector space representations of structured data such as XML and other
textual features such as phrases. As with many popular machine learning
approaches, most increases in quality with respect to human judges come from
how the data is represented.

\subsection{Choice of Sparsity Parameters}
During our experiments we found that setting the sparsity of the random codes
for each term to 1 in 12 set to +1 and 1 in 12 to -1 worked most effectively.
As the density of the random codes or index vectors increases, the potential
for cross talk between the codes increases. When the sparsity is decreased too
far there is not enough information for the query to successfully match
against. There is an optimal point for sparsity with respect to a given set of
queries. Detailed analysis of the effect of sparsity, including automated
methods to learn the optimal sparsity are outside the scope of this paper and
will be investigated in further research.

\subsection{Query Signatures}
In order to search the collection with a given query, we need to generate a
query signature.  Query document vectors are generated using standard TF-IDF
weighting. This real valued query vector is then converted to a signature using
exactly the same process as used with document signatures.  All the weighted
query term signatures are added to create a real valued randomly projected
document. The sign-bits are then taken to form the binary signature.  It is of
course necessary to use exactly the same process and parameters in generating
the query signature as when generating document signatures.  

The use of term weights in generating the query ensures that query terms that
are a-priori more significant (as determined through TF-IDF or some other
weighting function) will tend to dominate the signature bits where there is a
collision \em and \em a conflict.  Of course there is no need for concern when
the two terms agree on the sign when there is a collision.  This is easily
understood by looking at a case where we have two query terms, for instance,
``space" and ``shuttle".  If the term ``shuttle" has a larger TF-IDF value then
for any bit position where the two terms disagree on the sign, the term
``shuttle" will dominate the sign in the signature.  When there are multiple
terms we effectively get a vote.

Document signatures are represented in binary form, where 1-bits correspond to
+1, and 0-bits correspond to -1.  Query signatures, before taking the sign bits,
may contain a mix of 3 classes of values: positive, negative, or zero.  This
depends on the signs of term signatures, and a value of zero is obtained when
none of the query terms occupy some bit positions.  As a matter of fact, with
short queries and sparse term signatures this is almost invariably the case.
These zero valued bit positions are those for which the query does not specify
any preference. To account for this, a query mask is also generated to accompany
the query signature. This mask has 1-bits in all positions other than those that
are not covered by any term in the query.  The set bits in the mask identify the
subspace in signature space which the query terms cover.  When comparing the
query signature against document signatures, the similarity measure must not
take account of differences in those bit positions.  Conceptually, those are
neither $+1$ nor $-1$.
 
\subsection{Ranking}
\label{sec:ranking}
Ranking with TopSig is performed with the Hamming distance, calculating the
similarity score for each document.  The Hamming distance is rank equivalent to
the Euclidean distance since all signatures have the same vector length -- we
note that the signatures correspond to $+1$ and $-1$ values, not $1$ and $0$
values, and hence the length of each signature of $N$ bits is $\sqrt{N}$.  Since
the mask is almost invariably different for each query, the Hamming distance for
each query will generally be calculated in a different subspace.  The distance
metric is therefore a \em{masked} \em Hamming Distance.

If the document and query are identical in the query subspace then the Hamming
distance will be zero. The Hamming distance between two signatures of $N$ bits
is restricted to the range $[0,1,2,..N]$.  For a signature file with 1024 bits
per document there are at most 1025 possible distances between the query and a
document, and many less if the query is short. This means that in a collection
such as the Wikipedia, with millions of documents, if we rank all the documents
by the Hamming distance from the query, we are bound to get numerous ties.

Although document signatures are not completely random -- they are biased by the
document contents, and similar documents have similar signatures -- we still
expect the vast majority of the documents to be centred at about a Hamming
distance of $N/2$ from the query signature.  Indeed this is always observed.
The distribution of distances always resembles a binomial distribution, which we
expect if the distribution of signatures was indeed random.  It is not quite
that, but we still observe strong resemblance to truly random distribution.

We are interested in early precision and so TopSig can still achieve granularity
in ranking of documents.  This is because a large number of documents fall
much closer than $N/2$ to the query signature, and the number of ties diminishes
rapidly as the distance becomes smaller.    Some ties still remain nevertheless
and these may be broken arbitrarily or by using simple heuristics or document
features.  For instance, page-rank can be used, or any one of hundreds of
document features that are reportedly used in commercial search engines.

\subsubsection{Partial Index Scanning}
Given an index where each document signature is $N$ bits wide it is possible
scan only the first $f$ bits of each signature. This allows for further
decreases in time taken to rank. A multiple pass approach is possible where
the documents are first ranked with relatively few bits such as 640. The top
ten percent of the documents ranked using 640-bits can then be re-ranked using
the full precision of the document signatures stored in the index.

\subsection{Relevance Feedback}
Pseudo relevance feedback is known to improve the performance of a retrieval
system. TopSig can implement pseudo-relevance in the usual manner, through query
expansion. This however is a generic approach and can be used with any search
engine.  There is however an additional opportunity to apply pseudo-relevance
feedback, an opportunity that is unique and specific to TopSig. Explanation of
pseudo-relevance feedback is required to completely describe the approach we
have taken to ad-hoc retrieval with TopSig.

An initial TopSig search is first executed in the manner previously described
in Section \ref{sec:ranking}.
This search is performed in the subspace of the query signature, the subspace
spanned by the query terms.  This is achieved by using the masked Hamming
distance to rank all the documents in the collection.  Now it is possible to
proceed and apply pseudo relevance.  The principle is the same as with all
pseudo relevance approaches -- use some of the top ranked results to inform a
subsequent search.

We take the \em{top-k} \em ranked documents and create a new  query signature by
computing the arithmetic mean of the corresponding signatures by treating the
signatures as integer valued vectors and then taking the sign-bits in the
manner described in Section \ref{sec:packing}.  Since this signature was is
generated from full document signatures, this signature is now spanning the full
signature space and takes into account information from highly ranked results,
including in bit positions that were not informed directly by the query terms.
Now the query signature is in fact based on the full content of the nearest $k$
documents, through their signatures. The new query is constructed by inserting
only the missing bits into the original signature.  Therefore, the new signature
consists of the original signature in all originally unmasked positions, and the
feedback signature in all previously masked positions.  

The ranking of documents in relation to the new query is then repeated, but it
is not necessary to search the entire collection again.  It is sufficient to
re-rank a very small fraction of the nearest signatures -- usually those that
were retained in a shortened result list following the initial search.  This
step is consuming a negligible amount of additional computation -- several
orders of magnitude less than the initial search.   The feedback leads to
statistically significant improvement in performance.

The approach to pseudo-relevance feedback we have described exploits the binary
representation used by TopSig. This is conceptually similar to standard
pseudo-relevance feedback where the goal is to learn meaningful weights for
relevant terms not in the query. However, the implementation of the approach
with TopSig is drastically different as we work directly in the dimensionality
reduced space of the binary document signatures, rather than with specific
terms not in the original query.

\section{Evaluation and Results}
\label{sec:eval_results}
We have evaluated TopSig using the INEX Wikipedia 2009 collection, and the TREC
Wall Street Journal (WSJ) Collection. INEX Wikipedia collection contains
2,666,192 documents with a vocabulary of 2,132,352 terms.  The mean document
length in the Wikipedia has 360 terms, the shortest has 1 term and the longest
has 38,740 terms.   We used all 68 queries from INEX 2009 for which there are
relevance judgments. The Wall Street Journal Collection consists of 173,252
documents and a vocabulary of 113,288 terms.  The mean WSJ document length is
475 terms, the shortest has 3 terms, and the longest has 12,811 terms. We used
TREC WSJ queries 51-100.  

To compare TopSig with state-of-the-art approaches, we have used the ATIRE
search engine \cite{Trotman2010} which was formerly known as ANT.  ATIRE is a
highly efficient state-of-the-art system which implements several ranking
functions, over an inverted file system. The ATIRE search engine has been
thoroughly tested at INEX against other search engines, including several well
known systems such as Zettair,  Lucene, and Indri, and has been shown to produce
accurate and reliable results.

The references given herein to the ranking functions that were compared with
TopSig, are to the actual papers that were followed in implementing the methods,
as documented in the ATIRE search engine manual.  These are Jelineck-Mercer
(LMJM) \cite{Zhai2004}, DLH13 \cite{Macdonald2006}, Divergence from Randomness
\cite{Amati2002}, and Bose-Einstein \cite{Amati2002}.   The ranking functions
were evaluated with relevance judgments from TREC and INEX, and the \em
trec\_eval \em program.

\subsection{Recall-Precision}
\begin{figure}
\includegraphics[width=\columnwidth]{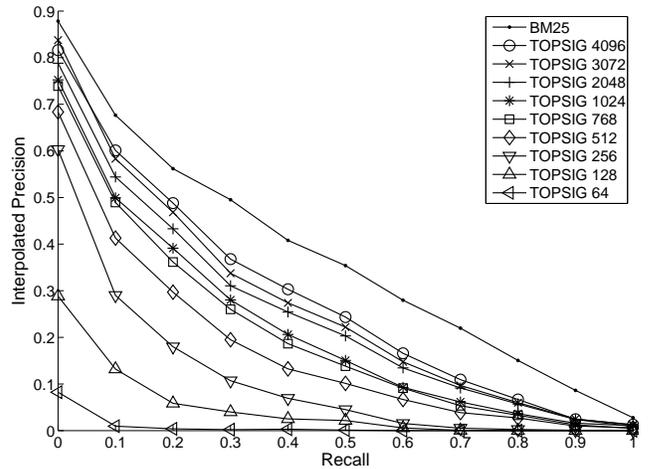}
\caption{INEX 2009 Precision vs Recall}
\label{fig:inex09pr}
\end{figure}

We first look at recall and precision over the full range of recall values.
Figure \ref{fig:inex09pr} depicts the precision-recall curves for INEX 2009
topics, against a tuned BM25 system, using $k=1.1$ and $b=3$, and with Rocchio
pseudo relevance feedback.   This BM25 baseline curve is an optimistic
over-fitted approach -- it is
tuned with the actual queries, and indeed performs better than any official run
at INEX 2009.  But we are concerned with evaluating TopSig and so this provides
very conservative yardstick by which to measure the performance.  The figure
shows several TopSig indexes, encoding the signature with 64, 128, 256, 512,
768, 1024, 2048, 3072, and 4096 bits per signature.  Only one in 12 vector
elements were set in the random term signatures, to either $+1$ or to $-1$, with
the rest of the elements set to $0$. It is interesting that
even a 64 bit
signature produced measurable early precision.  As the number of bits in the
signature increases, so does the recall.  The performance of the file signatures
is quite respectable once we allow for about 512 bits per signature --
particularly at early precision.

All the other language model based ranking functions produce a recall-precision
curve that falls below BM25, and just above the best TopSig curve, but are not
shown on the plot so as to reduce the clutter. Note that the legends in the
figures are ordered in decreasing order of area under the curve.

\subsection{Early Recall}
While Figure~\ref{fig:inex09pr}  may at first suggest that file signatures
produce inferior retrieval quality, we must focus our attention on the
early precision, and this requires some justification before we do that.

Moffat and Zobel \cite{Moffat2008} found that P@$n$ correlates with user
satisfaction. A user who is given 7 relevant documents in the top 10 is better
off than one who is only given 2. They argue that recall does not have a similar
use case that reflects user satisfaction. Even for a recall oriented task, a
user is unlikely to look past the top 30 results. For most tasks, the first page
or top 10 results are most useful to the user. Users achieve recall not through
searching the entire ranked list but by reformulating queries. Recent work by
Zobel and Moffat \cite{Zobel2009} suggests recall is not important except for a
few recall oriented tasks such as retrieval in medical and legal domains. If a
system provides 100\% recall, it implies that the user can create a perfect
query. Even in recall-based tasks, users tend to re-probe the collection with
multiple queries to minimise the risk they have missed important documents.

The same argument is applied to discount the importance that is attributed the
commonly used measure of Mean Average Precision (MAP) as it too depends on
higher recall and a long tail of relevant results. Again, it is not clear what
user satisfaction is correlated with MAP. Turpin and Scholer \cite{Turpin2006}
performed retrieval experiments where users completed search tasks using search
results with MAP scores between 55\% and 95\%. They were unable to find a
correlation between MAP scores and a precision based task requiring the first
relevant document to be found. For recall-based tasks, they only found a weak
link between MAP and the number of relevant documents found in a given time
period. They conclude that MAP does not correlate with user performance on
simple information finding web search tasks.

\subsection{Analysis of Early Recall}
Recall is not likely to be important to users except in some specific domains.
Therefore, we focus our attention on comparison of P@$n$ results between TopSig
and state of the art inverted file approaches.  The results immediately make it
obvious that TopSig is a viable option for common information finding tasks

\begin{figure}
\includegraphics[width=\columnwidth]{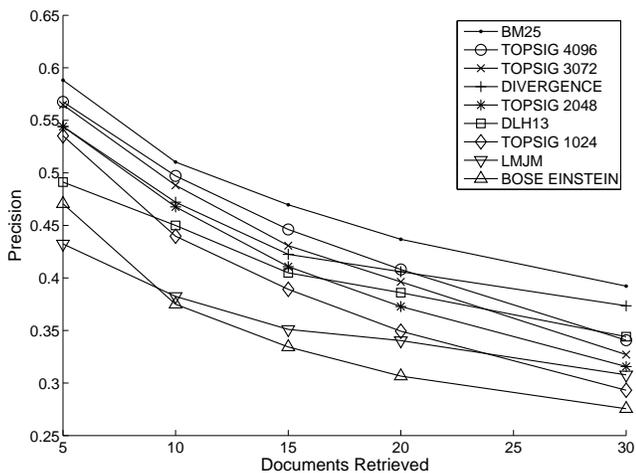}
\caption{INEX 2009 P@$n$}
\label{fig:inex09pat}
\end{figure}

\begin{figure}
\includegraphics[width=\columnwidth]{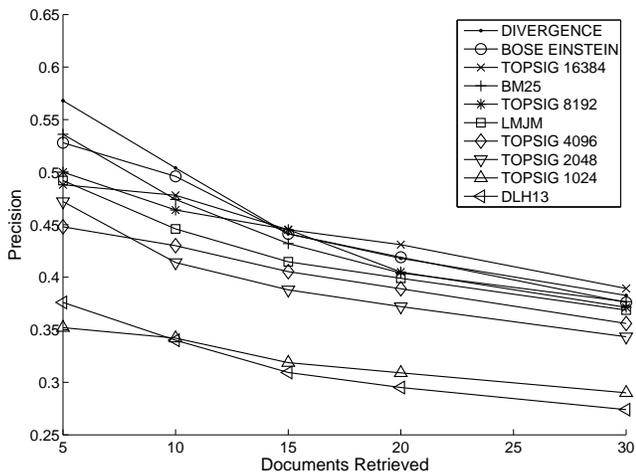}
\caption{Wall Street Journal P@$n$}
\label{fig:wsj_pat}
\end{figure}

To assess TopSig at early precision we look at early precision in the P@$n$
plots on Figures \ref{fig:inex09pat} and \ref{fig:wsj_pat}, for the 68 INEX 2009
ad-hoc queries and the TREC Wall Street Journal queries 51-100.  It is
immediately clear that TopSig performs similarly. The {\em only} system that
consistently outperforms TopSig is the over-fitted BM25 baseline. The legends
in the figures are ordered by the area under the curve, so that the best
performing systems appear first in the legend.

\begin{figure}
\includegraphics[width=\columnwidth]{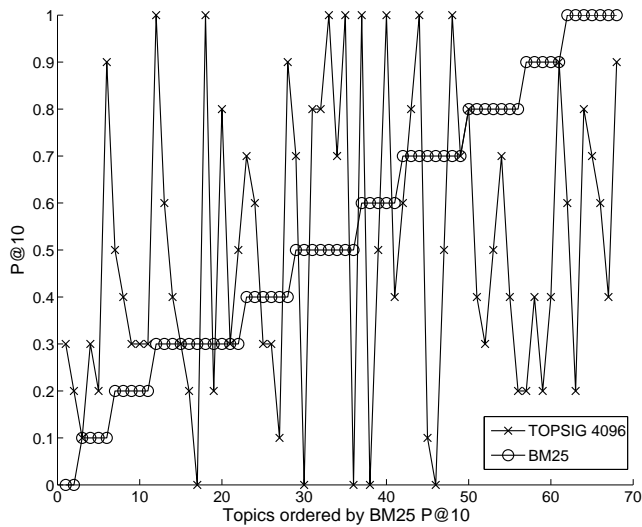}
\caption{INEX 2009 P@10 by Topic}
\label{fig:pat10_inex_bytopic}
\end{figure}

 In order to look more carefully at the differences, we focused on the P@10
performance differences on the INEX Wikipedia collection, between the best
performing ranking function -- BM25, and TopSig with 4096 bit signatures.  The
average P@10 for BM25 is $0.54$, and for TopSig it is $0.51$.  We look at all 68
queries and performed two-tailed paired $t$-test.  There is no statistically
significant difference with $p=0.41$. Figure ~\ref{fig:pat10_inex_bytopic}
depicts the P@10 values for all 68 queries.  The topics on the X-axis are
ordered by increasing P@10 values for BM25.  The TopSig P@10 values are plotted
in the same order.   It is obvious that the two approaches produce very
different results on a per-topic comparison.  The two systems do not agree on
which topics are difficult and which are not, and both sometime fail (on
different topics) to produce any relevant result in the top-10.  It is a common
and well understood phenomena that this should occur and it is true for all the
ranking functions that we tested.  However, there is a much stronger correlation
between all the language models, and BM25, as to which topics are hard and which
are easy.  No such correlation is observed for TopSig which seems to behave
quite differently despite producing similar overall precision. This leads us to
conjecture that combining TopSig with BM25 (or any of the other models tested)
may lead to better results than emerge from combining any other pair of more
correlated ranking functions.  Testing this conjecture is outside the scope of
this paper.

By inspecting at Figures \ref{fig:inex09pr} through \ref{fig:wsj_pat} one can
observe that as the number of bits in a document signature increases, the
quality of the results increases logarithmically. As more and more bits are
added the increases in quality become smaller and smaller. This agrees with the
Johnson-Lindenstrauss lemma \cite{Johnson1984} that states that the number of
dimensions needed to embed a high dimensional Euclidean space into one of much
lower dimension is logarithmic in the number of points.

\subsection{Storage and Processing overheads}
TopSig is efficient and compares well with the inverted file approach.  On a
standard PC, a $1024$ bit signature index can be searched by brute force in
about 175 milliseconds, with a collection of 2.7 million signatures of the
English Wikipedia documents.  The signatures file size for this collection is
only 325 MB, less than 0.65\% of the collection size, and so it easily fits in
memory.  By contrast, the highly compressed inverted file of ATIRE that
underlies all the other models, occupies 1.5GB, or about 3\% of the uncompressed
text collection size.  ATIRE itself is highly efficient and for comparison, the
Indri index for the same collection occupies about 11\% of the space.

Searching with TopSig is also efficient.  We have not implemented a parallel
multi-processor search which offers linear speedup in the number of CPUs.  Even
so,
all 68 queries for INEX collection were completed in 12 seconds for the
Wikipedia collection and all 50 WSJ topics were completed in 4 seconds, on a
basic Laptop.  This is comparable to the performance that is obtained with the
inverted file system.

There is potential to further compress the index by sorting the binary strings
lexicographically. Huffman coding can be used after sorting to represent the
differences between successive document signatures. This approach has been shown
to reduce a similar index used for near duplicate detection by up to 50\% when
used with 64-bit codes \cite{Manku2007}. Thorough testing of this style of
approach is beyond the scope of this paper and is expected to be investigated in
further research. It is also possible that document clustering can provide
effective ways to further compress the index. Document signatures in a
cluster fall within a small Hamming distance of each other. Therefore, only a
few bits differ between the cluster representative and document signatures it
represents.

\section{Clustering Evaluation}
\label{sec:clustering}
The goal of clustering is to place documents into topical groups. To achieve
this, clustering algorithms compare similarity between entire document vectors.
Therefore, the space and time efficiency of the TopSig representation allows it
to outperform current approaches using sparse vector representations. It is also
competitive in terms of document cluster quality. We have modified the k-means
algorithm to work with signatures. This approach is compared to the
implementation of k-means in the CLUTO clustering toolkit \cite{Karypis2002}
that is popular in the IR community. CLUTO uses full precision sparse vectors to
represent documents.

The same approach is used to create document signatures as for ad-hoc retrieval
as described in Section~\ref{sec:adhoc}. The sparsity of the signatures does not
have a large impact on the cluster quality but we found that index vectors with
1 in 6 bits set performed best. Index vectors with 1 in 3 and 1 in 12 bits set
were also tested.

The k-means algorithm \cite{Lloyd1982} was modified to work with the bit string
representation of TopSig. Cluster centroids and documents are $N$ bit strings.
Each bit in a centroid is the median for all documents it represents. If more
than half the documents contain a bit set to 1 then the centroid contains this
value in the corresponding position. As the 1 and 0 values represent +1 and -1,
this is equivalent to adding all the vectors and taking the sign of each
position. The standard Hamming distance measure is used to compare all vectors.
The algorithm is initialized by selecting $k$ random documents as centroids.
This modified version of k-means always converged when the maximum number of
iterations was not limited. Whether this modified version has the same
convergence guarantees as the original algorithm is unknown.

An implementation of the k-means clustering algorithm using bit-vectors is
available from the K-tree project subversion
repository \footnote{\url{
http://ktree.svn.sourceforge.net/viewvc/ktree/trunk/java/ktree/}} . Note that
this is an unoptimised Java implementation. It is expected further performance
increases can be gained by implementation in a lower level language such as C.

\subsection{Results}
We have evaluated document clustering using the INEX 2010 XML Mining collection
\cite{DeVries2011}. It is a 144,265 document subset of the INEX XML Wikipedia
collection. Clusters are evaluated using two approaches. The standard approach
of comparing clusters to a ``ground truth'' set of categories is measured via
Micro Purity. Purity is the proportion of a cluster that is the majority
category label. The final score is Micro averaged where the Purity for each
cluster is weighted by its size. On this collection, Purity produces
approximately the same relationships between different clustering approaches as
F1, Normalized Mutual Information and Entropy. There are 36 categories for
documents that are extracted from the Wikipedia category graph. 

An alternative evaluation is performed that has a specific application in
information retrieval. Ad-hoc retrieval relevance judgments are used to measure
the spread of relevant documents over clusters. This is motivated by the cluster
hypothesis \cite{Rijsbergen1979}, stating that documents relevant to the same
information need tend to cluster together. If this hypothesis holds then most of
the results will be in a small number of clusters. The Normalized Cumulative
Cluster Gain measure represents how relevant documents are spread over clusters.
It falls in the range $[0,1]$ where a score of 1 indicates all relevant
documents were contained in 1 cluster and a score of 0 indicates all relevant
documents were evenly spread across all clusters. Complete details of the
evaluation are available in a track overview paper from INEX 2010
\cite{DeVries2011}.

The sparse document vectors used to create the TopSig document signatures are
used as input to the k-means implementation in CLUTO. Therefore, we are
comparing the same algorithm on the same data except for the fact the TopSig
representation is extremely compressed and has a different centroid
representation and distance measure. Both implementations of k-means are
initialized randomly and are allowed to run for a maximum of 10 iterations. 36,
100, 200, 500 and 1000 clusters were produced by each approach where 36 was
chosen to match the number of categories. This allows the trend of the measures
to be visualised as the number of clusters are varied.

\subsection{Analysis of Results}
\begin{figure}
\includegraphics[width=\columnwidth]{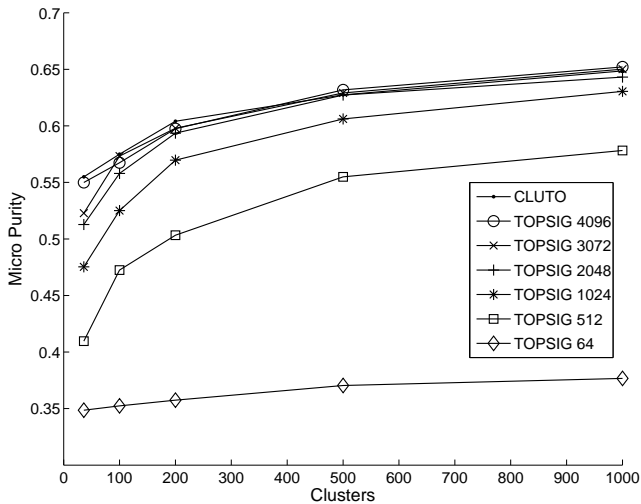}
\caption{INEX 2010 Micro Purity}
\label{fig:inex10purity}
\end{figure}

\begin{figure}
\includegraphics[width=\columnwidth]{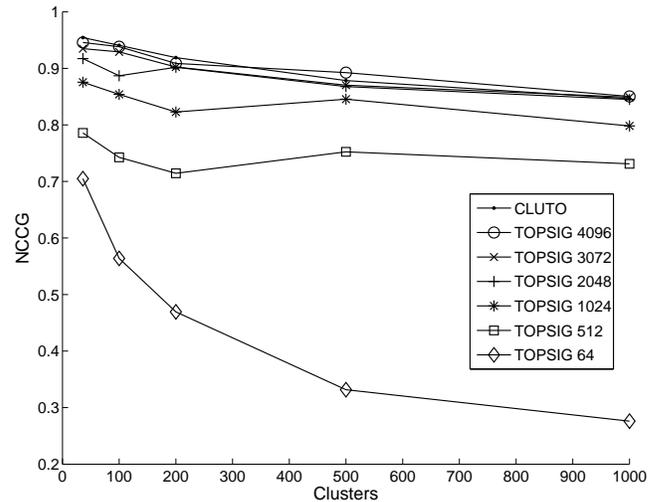}
\caption{INEX 2010 NCCG}
\label{fig:inex10nccg}
\end{figure}

Figures~\ref{fig:inex10purity}~and~\ref{fig:inex10nccg} represent the quality
of the clustering approaches using the Micro Purity and NCCG measures
respectively. The TopSig representation nears the quality of CLUTO at 1024 bits
and matches it at 4096 bits according to both measures. The best NCCG scores are
all greater than 0.84 for all numbers of clusters, strongly supporting the
cluster hypothesis, even when splitting the collection into 1,000 clusters.

\begin{figure}
\includegraphics[width=\columnwidth]{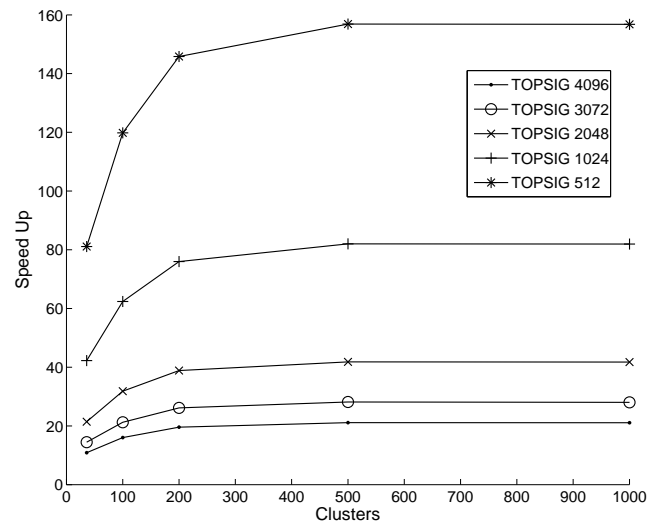}
\caption{INEX 2010 Clustering Speed Up}
\label{fig:inex10speedup}
\end{figure}

Figure~\ref{fig:inex10speedup} shows how many times faster the TopSig
clustering is than the traditional sparse vector approach in CLUTO. For example,
using 4096 bit signatures to create 500 clusters is completed 20 times faster
than CLUTO and 80 times faster at 1024 bits. This is one to two orders of
magnitude increase in efficiency while still achieving the same quality as
traditional approaches.

Figures~\ref{fig:inex10purity},~\ref{fig:inex10nccg}~and~\ref{fig:inex10speedup}
can not be significance tested as they are a single run of the algorithms.
However, the graphs allow the general trends to be visualised. CLUTO k-means
takes approximately 5 hours to produce 1,000 clusters on this relatively small
collection. Therefore, the CLUTO and TopSig k-means algorithms were repeatedly
run to produce 36 clusters given different starting conditions. Given each
random initialisation, k-means converges to a different local minima. The
k-means implementations were run 20 times to measure this variability.
Table~\ref{table:36clusters} contains the results of this experiment where
TopSig approaches that are equivalent to the CLUTO approach are highlighted in
boldface. Equivalence was tested using the $t$-test with $p$ > 0.05 indicating
no statistically significant difference.

The time to produce the document signatures from the sparse document
vectors was not included in the evaluation. The time is negligible in
comparison to the time it takes to cluster using sparse document
representations. Furthermore, when the k-means algorithm is limited in the
number of iterations it can run for, it's complexity becomes linear. The
complexity of the document signature generation is also linear in the number of
non-zero ($nnz$) elements in the term-by-document matrix. As, $O(nnz) + O(nnz)
= O(nnz)$, the complexity of the clustering system is not changed by the
introduction of the generation of the document signatures.

\begin{table}
\begin{center}
\begin{tabular}{lccc}
\hline\noalign{\smallskip}
Method & Micro Purity & NCCG \\
\noalign{\smallskip}
\hline
\noalign{\smallskip}
CLUTO & $0.543 \pm 0.008$ & $0.955 \pm 0.003$ \\
TopSig 4096 & ${\bf 0.540 \pm 0.008}$ & ${\bf 0.951 \pm 0.007}$ \\
TopSig 3072 & $0.528 \pm 0.009$ & $0.939 \pm 0.005$ \\
TopSig 2048 & $0.520 \pm 0.007$ & $0.926 \pm 0.007$ \\
TopSig 1024 & $0.480 \pm 0.007$ & $0.867 \pm 0.012$ \\
\noalign{\smallskip}
\hline
\noalign{\smallskip}
\end{tabular}
\caption{Detailed Evaluation of 36 clusters}
\label{table:36clusters}
\end{center}
\end{table}

\newpage
\section{Discussion}
\label{sec:discussion}
We have described TopSig, an approach to the construction of file signatures
that emerges from aggressive compression of the conventional term-by-document
weight matrix that underlies the most common and most successful inverted file
approaches.  When focusing on early precision, using P@$n$ measures from P@5 up
to
P@30, TopSig is shown to be as effective as even the best models available,
while requiring equal or less amounts of space for storing signatures.
Significant reductions in signature index size can be achieved with TopSig as a
trade-off, reducing the signature file size by orders of magnitude, while
accepting reduced early precision.  Remarkably, even a single double precision
variable -- a 64bit signature -- is found to achieve 10\% P@10 over 2.7 million
documents Wikipedia collection.  Testing with standard clustering benchmark
tasks demonstrates TopSig to be equally effective and as accurate as a
state-of-the-art clustering solution such as CLUTO, with processing speedup of
one to two orders of magnitude.

TopSig had been applied to documents of greatly varying lengths.  Both the WSJ
and the Wikipedia collections have very short to very long documents, varying in
size by up to five orders of magnitude.  It had been suggested that file
signatures are susceptible to this situation because of the increased
probability of collisions on terms, but TopSig still performs well on these
collections. In particular, we have tested TopSig with WSJ -- the same
collection that was used by Zobel et al to demonstrate the superiority of
conventional inverted files. TopSig clearly outperforms conventional file
signatures that were previously discredited.  In this paper we compare TopSig
directly with inverted file approaches to demonstrate similar performance
levels.

Unlike early approaches to searching with file signatures, TopSig does not
necessitate the complicated and tedious removal of false matches, and supports
ranked retrieval in a straight forward manner.  All the performance evaluation
results that are reported in this paper were performed without any attention
being paid to false matches.  Not only is TopSig producing comparable results,
but with respect to false matches it is also virtually indistinguishable from a
user perspective because false matches do not occur unless using far  too
aggressive compression is applied, for instance, compressing documents into 64
bit signatures.

There are certain differences between TopSig and inverted file based retrieval
which may offer advantages in some application settings.   TopSig performs the
search in constant time and independently of query length.  Comparing full
documents to the collection in a filtering task, or processing long queries,
take exactly the same time as comparing a single term query.  This may be useful
in applications where predictability and quality of service guarantees are
critical.  Shortening the signature length can reduce the index size,
with smooth degradation in retrieval performance.  Signatures may offer
significant advantages where storage space is at a premium and a robust
trade-off is sought.

Distributed search is an attractive setting for TopSig -- distributed indexing
and retrieval have to resolve the problems of collection splitting and result
fusion.  With TopSig these operations are trivial to implement since the Hamming
Distance between signatures can be used as a universal metric across the system.
Gathering of global statistics can be ignored by using the raw term
frequencies from each document. This further simplifies use of TopSig in a
distributed setting and the trade-off with quality may be acceptable depending
on the particular use of the system. If each text object in an enterprise
carries its own signature -- perhaps
generated independently as a matter of routine by the applications that maintain
the objects -- then crawling and indexing the enterprise collection is a simple
as collecting the signatures. Alternatively, TopSig can support the
implementation of massively parallel search simply by distributing the query
signature to every participating sub-system that maintains its own set of
signatures.  It is also trivial to implement distributed filtering with TopSig
by maintaining a ``watch list'' of signatures that can be compared with incoming
text objects at run time.   TopSig is trivial to distribute on multi-processor
platforms for the very same reasons.  The simplicity of the search process means
that with shared memory processor architecture a linear speedup in the number of
concurrent hardware threads available can be achieved.

TopSig is particularly efficient in indexing.  It places virtually no memory
requirements during indexing, processing an entire collection in a single pass
(assuming term statistics are stable, which they are in very large collections).
 The most significant remaining drawback to TopSig is that it still requires a
comparison with all signatures in the collection.  Parallel processing offers a
simple solution, but it is not entirely satisfactory.  Parallel search does not
reduce the amount of computation that is required, it only distributes it.
There are many reports in the research literature about more efficient
approaches to signature file searching, which operate in sub-linear time.  Many
tree based approaches have been described, and some solutions offer
improvements.  It is not a solved problem by any means it is the subject of
ongoing research with TopSig too.  

\label{sec:conclusion}
This paper introduces TopSig, a new file signature approach that represents a
viable alternative to conventional search engines.   Our results demonstrate
that with a different approach to signature construction and searching file
signatures performance is comparable to that of conventional language and
probabilistic models at early precision.  TopSig represents a principled
approach to the construction of file signatures, placing it in the same
conceptual framework as other models.  This is very different from the
conventional ad-hoc formulation of file signatures.  
Future work with TopSig will address multi-processor implementation, a tree
structured approach to the search process, and evaluation in a massively
parallel massively distributed setting.  Early findings of experiments with
longer documents indicate that even improved performance can be achieved with
TopSig by splitting documents.  This  is the subject of ongoing research.

\bibliographystyle{plain}
\bibliography{bib}

\end{document}